\documentclass[manuscript,natbib209, preprint]{aastex}
\usepackage{amssymb, amsmath, fullpage, graphicx, bm, natbib}

\shorttitle{The early history of protostellar disks, outflows, and binary stars}
\shortauthors{Duffin \& Pudritz}

\begin{document}

\title{The early history of protostellar disks, outflows and binary stars}
\author{Dennis F. Duffin}
\affil{Department of Physics and Astronomy, McMaster University, Hamilton ON L8S 4M1, Canada}
\email{duffindf@mcmaster.ca}

\and

\author{Ralph E. Pudritz}
\affil{Department of Physics and Astronomy, McMaster University, Hamilton ON L8S 4M1, Canada}
\affil{Origins Institute, McMaster University, Hamilton ON L8S 4M1, Canada}
\email{pudritz@physics.mcmaster.ca}

\begin{abstract}
In star formation, magnetic fields act as a cosmic angular momentum extractor that increases mass accretion rates onto protostars and in the process, creates spectacular outflows. However, recently it has been argued that this magnetic brake is so strong that early protostellar disks -- the cradles of planet formation -- cannot form. Our three-dimensional numerical simulations of the early stages of collapse ($\lesssim 10^5$ yr) of overdense star--forming clouds form early outflows and have magnetically regulated and rotationally dominated disks (inside 10 AU) with high accretion rates, despite the slip of the field through the mostly neutral gas. We find that in three dimensions, magnetic fields suppress gravitationally driven instabilities which would otherwise prevent young, well ordered disks from forming. Our simulations  have surprising consequences for the early formation of disks, their density and temperature structure, the mechanism and structure of early outflows, the flash heating of dust grains through ambipolar diffusion, and the origin of planets and binary stars.
\end{abstract}

%
%
\section{Introduction}
Over the past decade numerical simulations have enabled the exploration of the central
physical questions regarding the nature of star formation -- from the collapse of an initial dense gaseous molecular cloud 
core to the formation of a star and its associated protostellar disk and outflow, and the emergence of  a star's planetary
system.  The formation of disks and jets during star formation is central to many of these issues, but very little is known
about the earliest phases of their evolution.  How is 
the initial excessive angular momentum associated with the star's natal core removed  -- through magnetic braking \citep{1994ApJ...432..720B} and then outflows \citep{2006ApJ...641..949B,2007prpl.conf..277P}, or by spiral density waves in 
disks \citep[e.g.~][]{2009arXiv0901.4325L}?  Do multiple stars form 
through gravitational fragmentation of cores or massive disks \citep{HW2004b,2007MNRAS.377...77P,2008A&A...477...25H}?   What is the significance
of outflows and jets that are launched before most of the mass has collapsed into the disk \citep{2003MNRAS.341.1360L, 2006ApJ...641..949B}  in a wide variety of young
stellar systems -- from brown dwarfs \citep{2005Natur.435..652W, 2009ApJ...699L.157M} to massive stars \citep{2007prpl.conf..245A, 2007ApJ...660..479B}?  

Extraction of angular momentum by magnetic fields that thread the collapsing gas may be too efficient, according to recent two-dimensional  \citep{2008ApJ...681.1356M} and three-dimensional axisymmetric \citep{2008A&A...477....9H} simulations, preventing the formation of rotationally dominated disks, even when the effects of imperfect coupling of the field with the gas are included \citep{2009ApJ...698..922M}. 
These results seemingly contradict the 
observations which clearly show that disks are present around most if not all young stars, even 
in environments in which the magnetic field is expected to be strong \citep{1995AJ....109.1846H}. 
\citet{2008A&A...477....9H} performed three-dimensional ideal MHD simulations of collapsing cores using a barotropic equation of state, concluding that no rotationally dominant structure is formed from 10 to 100 AU for highly magnetized cores.   \citet{2008ApJ...681.1356M} performed two-dimensional ideal MHD simulations on collapsing singular isothermal toroids using a barotropic equation of state and an inner boundary at 6.7~AU  (effectively a sink particle), finding that even moderately magnetized disks could not form.   Such two-dimensional  models impose a high degree of mathematical symmetry  and therefore miss the formation of bars and spiral waves in disks.  

We improve on previous results through this three-dimensional adaptive mesh refinement (AMR) investigation which includes a full treatment of the cooling \citep{2006MNRAS.373.1091B} and the finite coupling of the magnetic field 
to the pre--stellar gas \citep[ambipolar diffusion,][]{2008MNRAS.391.1659D}.  We find that ordered disk--like structures can emerge on scales $\lesssim 10$ AU at early times ($\lesssim 10^5$ yr) in magnetized systems. We have omitted a sink particle in this study as they are expected to affect the solution to within a few sink radii.  Without sinks simulations are indeed limited to early disks, although they offer a full solution to the region within 10 AU where the heated core forms.

%
%
\section{Methods} 
We model an isolated star forming core with a rotating Bonnor--Ebert (BE) sphere 
 \citep{1956MNRAS.116..351B,1955ZA.....37..217E} commonly observed in nature \citep{2001Natur.409..159A, 2003ApJ...586..286L, 2005ApJ...629..276T}, using the FLASH AMR code.  The refinement 
criteria resolves the Jeans' length ($\lambda_\mathrm{J}=\sqrt{\pi c_s^2/G\rho}$) by at least eight cells, limited by the ambipolar diffusion time step. The ambipolar diffusion time step is 
\begin{equation}
\tau_\mathrm{AD} = 0.5\frac{\Delta x^2}{\eta_\mathrm{AD}} = 0.5\frac{\mu_0 \gamma_\mathrm{AD} \rho_i\rho_n \Delta x^2}{1.4 B^2},
\end{equation}
where $\Delta x$ is the cell length, $\eta_\mathrm{AD}$ is the diffusivity, $\gamma_\mathrm{AD} = 3.28\times 10^13 \mathrm{~g^{-1}~ cm^3 ~s^{-1}}$ is the ion-neutral collisional coupling, $B$ is the magnetic field strength, and $\rho_i = 2.32\times 10^{-25} \left( \rho_n / 10^{-18} \right)^{0.5} \mathrm{~g~cm^{-3}}$ \citep[e.g.~][]{2008MNRAS.391.1659D}.

We construct an initial BE sphere with a radius of $R=6.9 r_0 = 6.9 c_{s_\mathrm{core}} / \sqrt{4\pi G \rho_c}$ (where the critical radius is $R_c =6.49 r_0$) and add a 10\% over--density 
on top with a 10\%, $m=2$ perturbation to break axisymmetry.  Such perturbations model the buffeting that observed star--forming cores
undergo from supersonic turbulent gas motions that prevail in the surrounding low density molecular gas 
(our perturbation accounts for a 0.04\% change in gravitational binding energy, so this is a rather quiescent core). Our BE spheres have a mass of $1.18 M_\odot$ and initial central densities of $\rho_c = 6.07\times 10^{-18}~\mathrm{g~cm^{-3}}$. The magnetic field is added to the core in the $z$ direction -- in agreement with recent observations \citep{2009ApJ...701.1044K} -- such that 
the ratio of the thermal to the magnetic energy in the gas, known as the plasma 
beta ($\beta = 2 c_s^2 / v_\mathrm{A}^2$ = 46.01) is constant, where 
$c_s=0.27~\mathrm{km~s^{-1}}$ is the core's sound speed and the Alfvén speed is 
$v_\mathrm{A} = B / \sqrt{4\pi\rho} \simeq 0.74 c_s/ \Gamma$ (the latter 
relation for critical BE spheres).  The mass-to-flux ratio $\Gamma$ 
is a measure of the ratio of the gravitational to the magnetic energy in gravitating objects.  If 
$\Gamma<1$, the magnetic field is strong enough to prevent gravitational collapse and the core is 
called sub-critical, otherwise it is called supercritical.  We set the 
mass-to-flux $\Gamma = 2\pi G^{1/2}\Sigma/B = 3.5$, where $\Sigma$ is the 
column density, stemming from observations of magnetic fields in early cores 
which indicate $\Gamma\approx 1-4$ \citep{1999ApJ...520..706C} and confirmed 
through simulations of cloud-scale turbulence \citep{2007MNRAS.382...73T}. 
The rotation values of our models were taken from previous simulations of early 
star--forming clusters \citep{2007MNRAS.382...73T}. We take the extremes and average values of rotation and interpret these as low, moderate and high rotating cores (Table \ref{tab:1}).   The moderate rotation model set has a ratio of rotational to gravitational binding energy of $\beta_\mathrm{rot}=0.046$,  in line 
with observations \citep{2003ApJ...586..286L}.

For each model set, we run a non--magnetic (hydrodynamic), a perfectly coupled 
(ideal magnetohydrodynamic) and more realistic, partially coupled (ambipolar diffusion) case for a total of nine simulations (summarized in Table \ref{tab:1}).

%
%
\section{Do disks form?}  

By the time the maximum surface 
densities in the moderate rotation model set reach a value of
$\Sigma_c=4.2\times 10^3~\mathrm{g~cm^{-2}}$, 0.1 $M_\odot$ is contained 
within 156 AU (about 10\% of the core's mass is contained within 0.003\% of 
its initial volume).  The mass of the protostar (taken to be all material inside 
1 AU or $10^{-3}~M_\odot$) is only 6 \% of the total mass in the disk (taken to be everything inside 10 AU or $10^{-2}~M_\odot$).  At this point, the non--magnetic, perfectly coupled and partially coupled clouds are 0.160, 0.194, and 0.175 million years old respectively (corresponding to 5.9, 7.2 and 6.5 free-fall times, where $t_\mathrm{ff} = 0.027$ Myr is estimated using $\rho_c$). 
 An extended dense region has taken shape, the gas has been heated, 
and in the cases where magnetism is present, disk winds have started. 

We present in Figure \ref{fig:1}, snapshots of the structure of the region at the 
center of the collapse, taken at this time for our three different cases.    
 Despite having identical initial conditions,
 the non--magnetized, moderate rotation case has formed two bars, driving 
 off--axis rotational modes which themselves create an off--axis bar.   The wobbling in the collapse occurs through accretion along shocks and density perturbations in the gas. This stands in stark contrast to the obvious disk--like structures and outflows 
present in both ambipolar diffusion (Figure 1(b) and ideal MHD cases (Figures 1(c) and (d)\footnote{We provide movies of Figures \ref{fig:1}(a)-(c) online illustrating the density and magnetic field line structure on all scales in the simulation.}.  
Strong oscillations of the disk radius were noted in the simulations of \citet{2007MNRAS.377...77P} which were also dominated by massive disks in their early phase. They also found that this behavior was suppressed by magnetic fields.

Most importantly, the magnetic disks, by preventing strong bars from forming, 
are rotationally dominated whereas the uncoupled disk is not! This is illustrated 
in Figure \ref{fig:2}(a), where we see that the ambipolar diffusion case has a 
larger rotationally dominated disk ($v_\phi>v_\mathrm{infall}$) than the perfectly 
coupled case.  We note also, by plotting $v_\phi/v_\mathrm{infall}$ of the most evolved perfectly coupled 
state of the collapse, that the rotational support is \emph{growing} with time 
while hydrodynamic rotational support becomes increasingly unstable. 
In Figure \ref{fig:2}(a) we also plot surface averaged plots of 
the ratio of the rotational velocity to the Keplerian velocity -- the rotational velocity of a gas 
in a stable orbit around an enclosed mass.  We find peak ratios of about 0.1--0.5 
in all cases of the moderate rotation model (in fact some pre--binned values exceed 1 in the ambipolar case).

In the limit of low rotation, hydrodynamic
spiral waves and bars in the purely hydrodynamic case are confined to a region of about 1 AU in radius. 
This has little effect on the braking of the cloud (Figure \ref{fig:2}(d) and rotationally dominated disks are allowed to form outside of 1 AU (Figure \ref{fig:2}(b).  The hydrodynamic case 
is much more spherical in nature while the magnetized cases remain fairly flat,
despite efficient magnetic braking.  In the limit of low rotation,
Figure 2(b) shows that the rotation in the hydrodynamic case is nearly Keplerian and the
magnetic case is much more sub--Keplerian,
whereas in the case of moderate rotation shown in Figure 2(a) the opposite is true.
Thus in the extreme low rotation limit where spiral modes and bars are strongly suppressed, early hydrodynamic disks do form. 

 Due to the bar, the distribution of specific angular momentum $j(R)$ (where $R$ is the cylindrical radius; Figure \ref{fig:2}(d) of the hydrodynamic case is comparable to the cases where magnetic braking is present.  In the low rotation model set, the bar is small and $j(R)$ does not compare as well to the magnetized cases.  The hydrodynamic and perfectly coupled cases of the moderate rotation model set 
have similar maximal accretion rates of  
$\dot{M}_\mathrm{accr}\approx4\times 10^{-4}~\mathrm{M_\odot~yr^{-1}}$,
while the ambipolar case reaches maximal accretion rates of only 
$\dot{M}_\mathrm{accr}\approx1\times 10^{-4}~\mathrm{M_\odot~yr^{-1}}$ due 
to weaker magnetic braking and a stable disk. 
In the low rotation model set, mass accretion rates are $\dot{M}_\mathrm{accr}\approx2\times 10^{-4}~\mathrm{M_\odot~yr^{-1}}$ for the magnetized cases and $\dot{M}_\mathrm{accr}\approx3\times 10^{-4}~\mathrm{M_\odot~yr^{-1}}$ for the hydrodynamic case.

In the high rotation model set however, no cases form rotationally dominated disks as all three cases have formed some sort of bar on large scales (Figure \ref{fig:3}).  Clearly, it is through the suppression of instabilities in the collapse that magnetic fields promote early disk formation.  Disk formation in systems without magnetic fields typically must wait until the mass of the central star dominates that of the surroundings sufficiently to suppress these instabilities.    

%
%
\section{Early outflows and magnetic fields}  

\begin{figure}
\plottwo{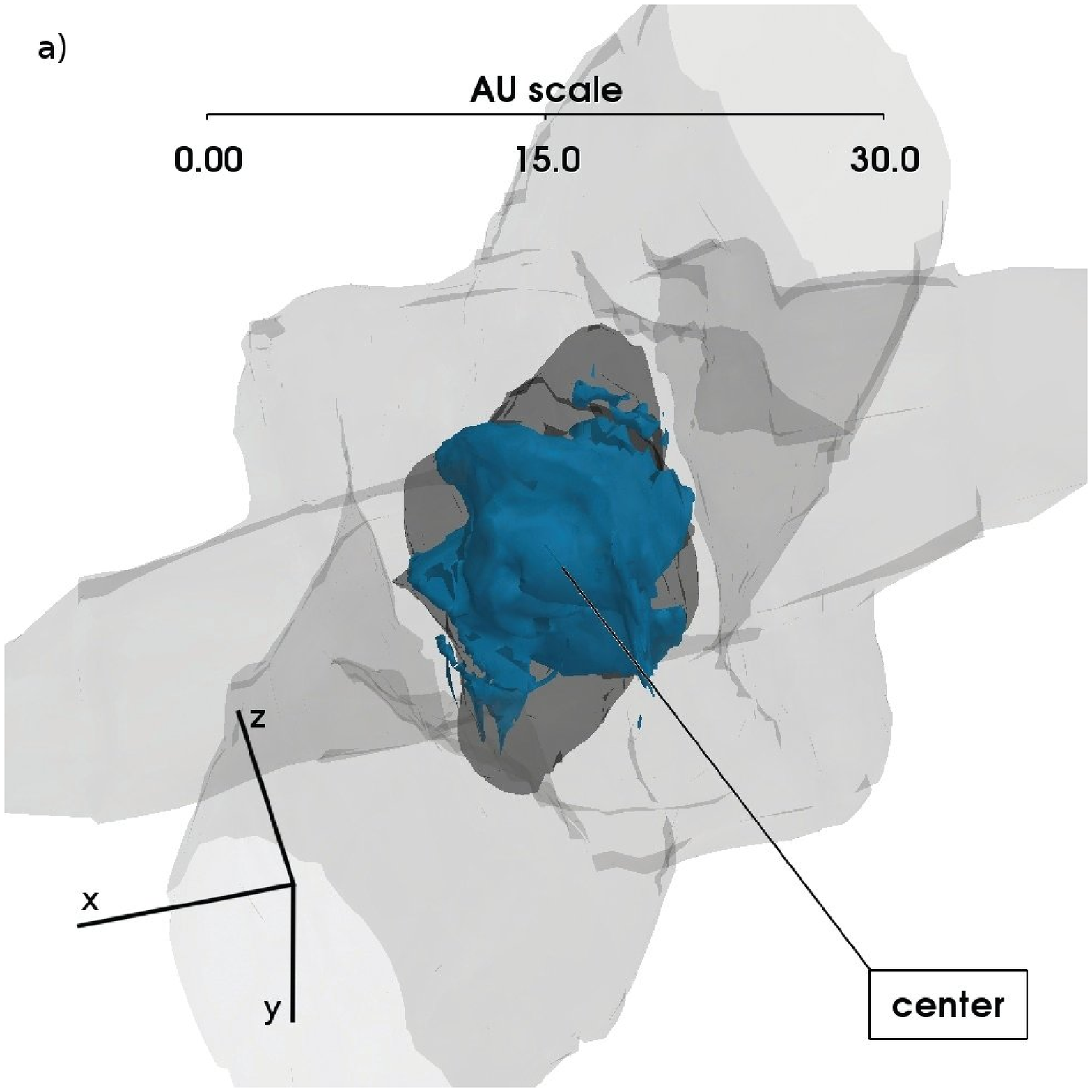}{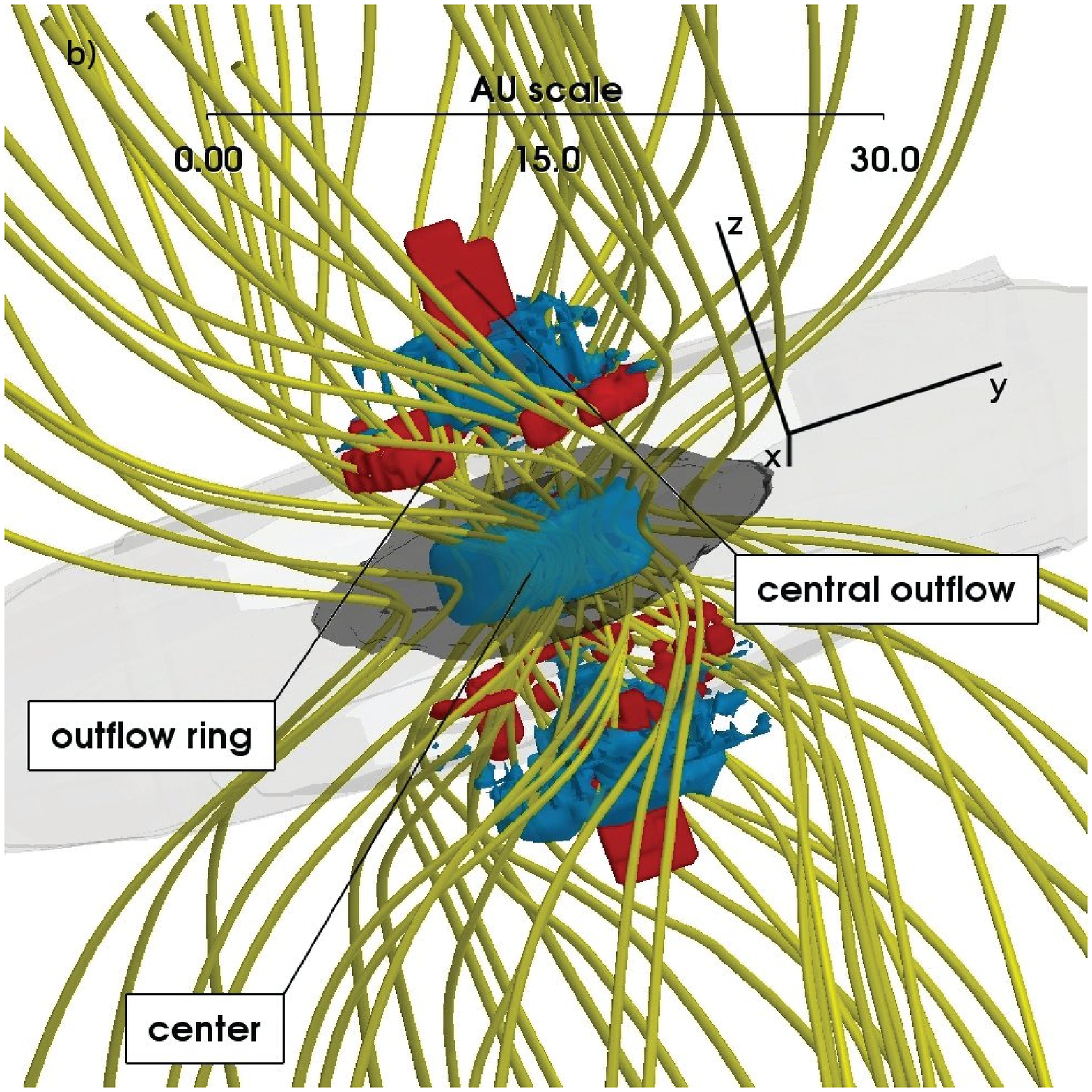}

\plottwo{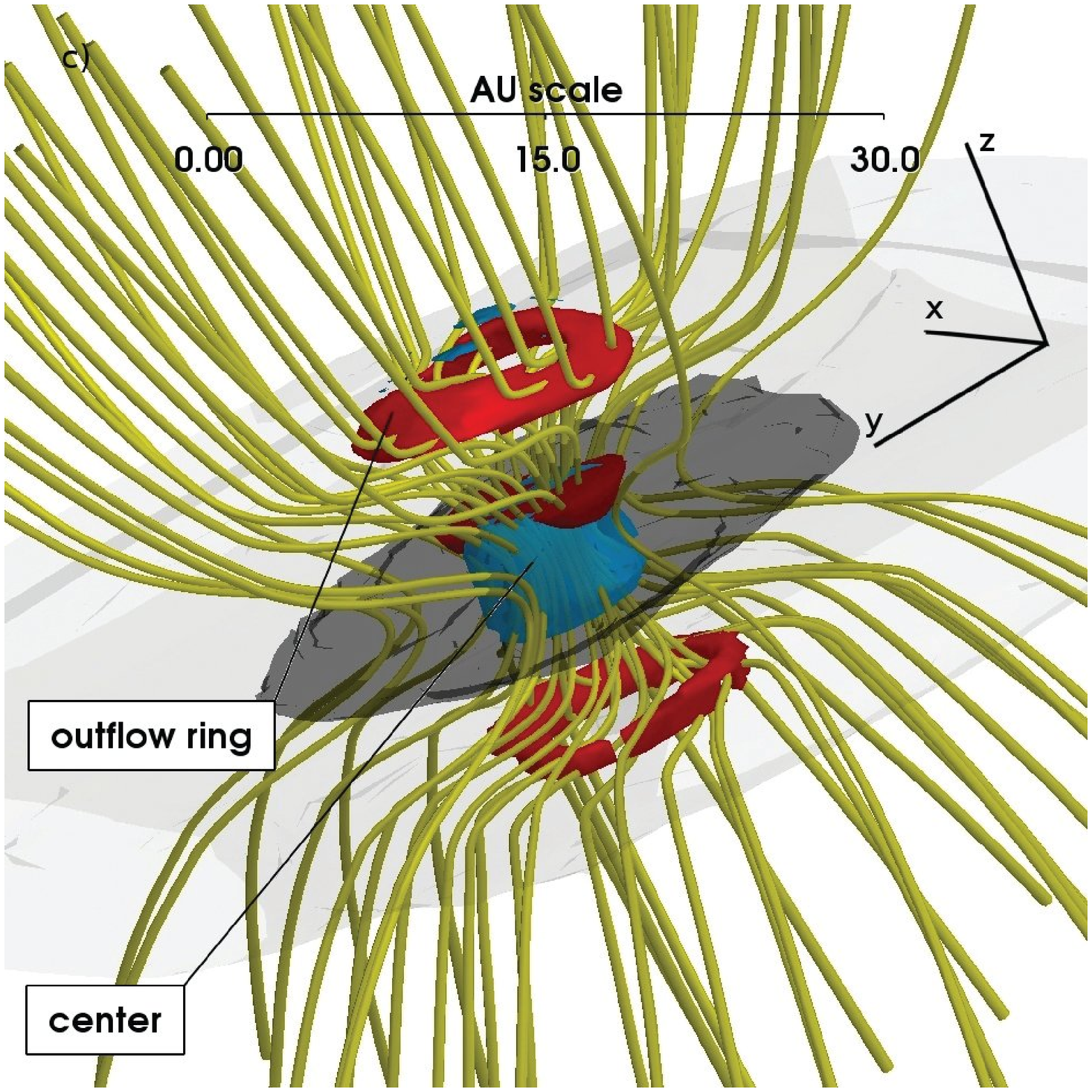}{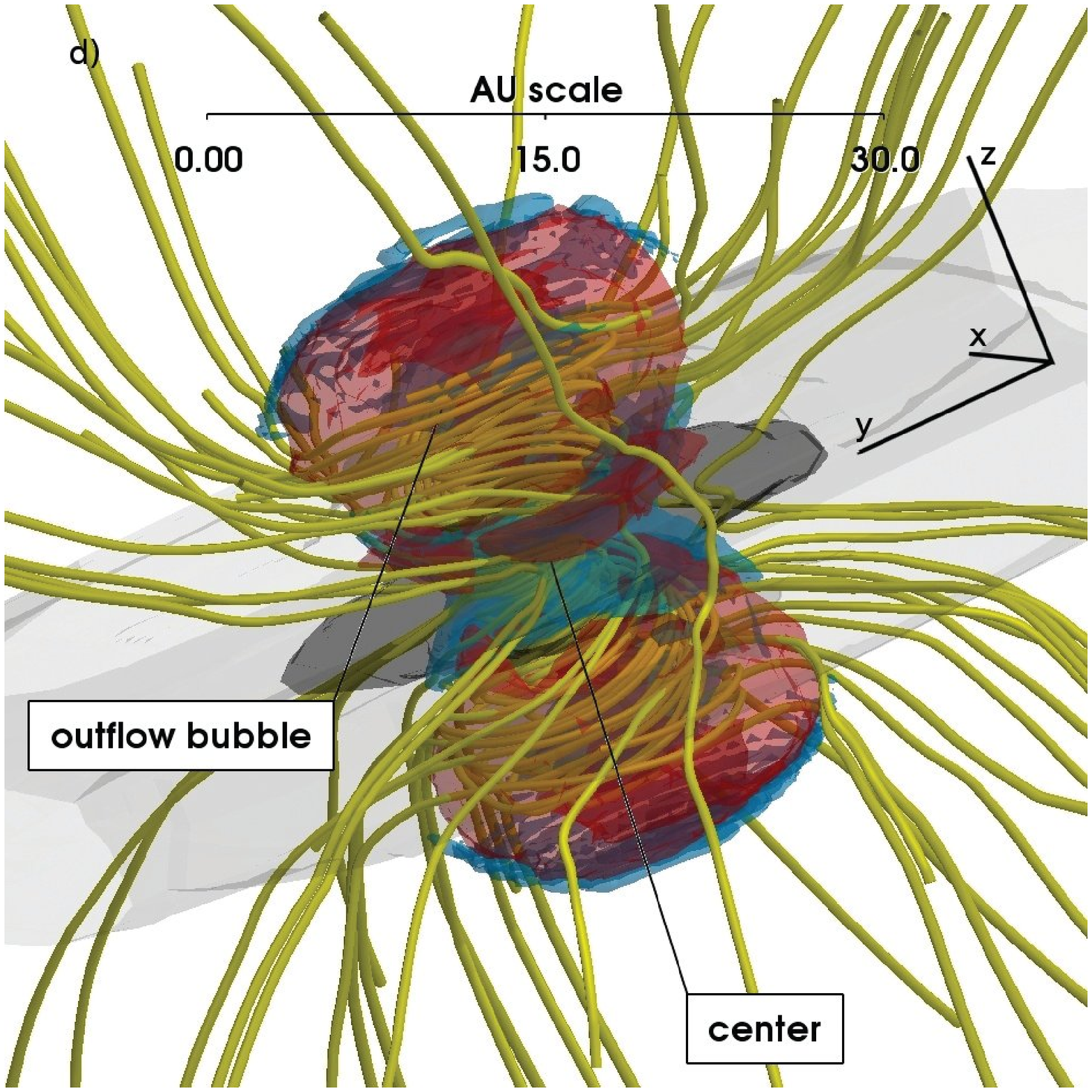}

\caption{\label{fig:1}
Inner disk of the moderate rotation model set at central surface densities of 
$\Sigma_0=4.2\times 10^3~\mathrm{g~cm^{-2}}$ for (a) hydrodynamic,
 (b) ambipolar diffusion, (c) perfectly coupled cases, and (d) the evolved state of the 
 perfectly coupled case at $\Sigma_c = 8.72\times10^7~\mathrm{g~cm^{-2}}$. 
 Black and gray surfaces are $\rho=10^{-12}~\mathrm{g~cm^{-3}}$ 
 ($n=2.58\times10^{11}~\mathrm{cm^{-3}}$) and $\rho=10^{-13}~\mathrm{g~cm^{-3}}$ 
 ($n=2.58\times10^{10}~\mathrm{cm^{-3}}$), respectively. Temperature contours at $T=85~\mathrm{K}$ are 
 shown in blue. Outflow velocities contours, shown in 
 red, are $0.3~\mathrm{km~s^{-1}}$ in (c) and (d) and $0.01-0.4~\mathrm{km~s^{-1}}$ in b.  Magnetic field lines are shown as 
 yellow tubes. (Animations of this figure are available in the online journal.)}
\end{figure}

\begin{figure}

\plottwo{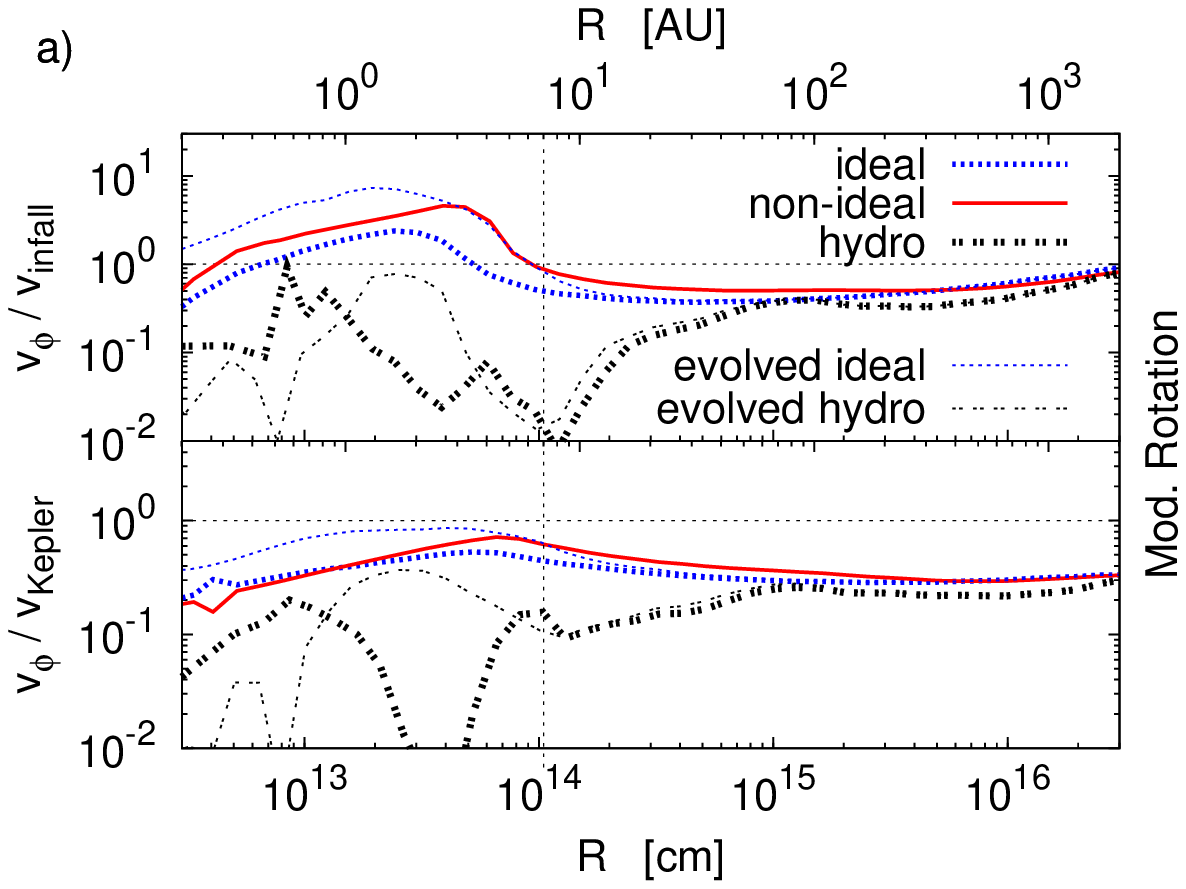}{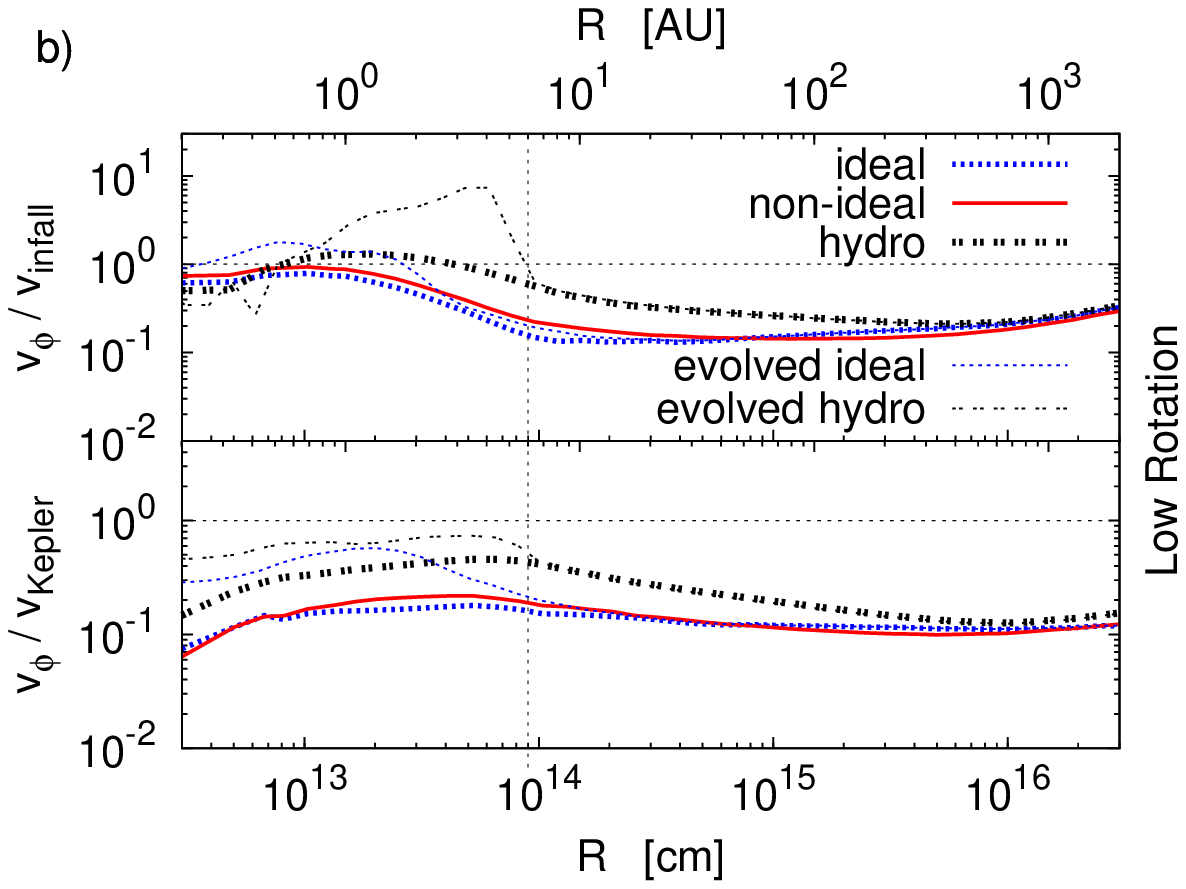}

\plottwo{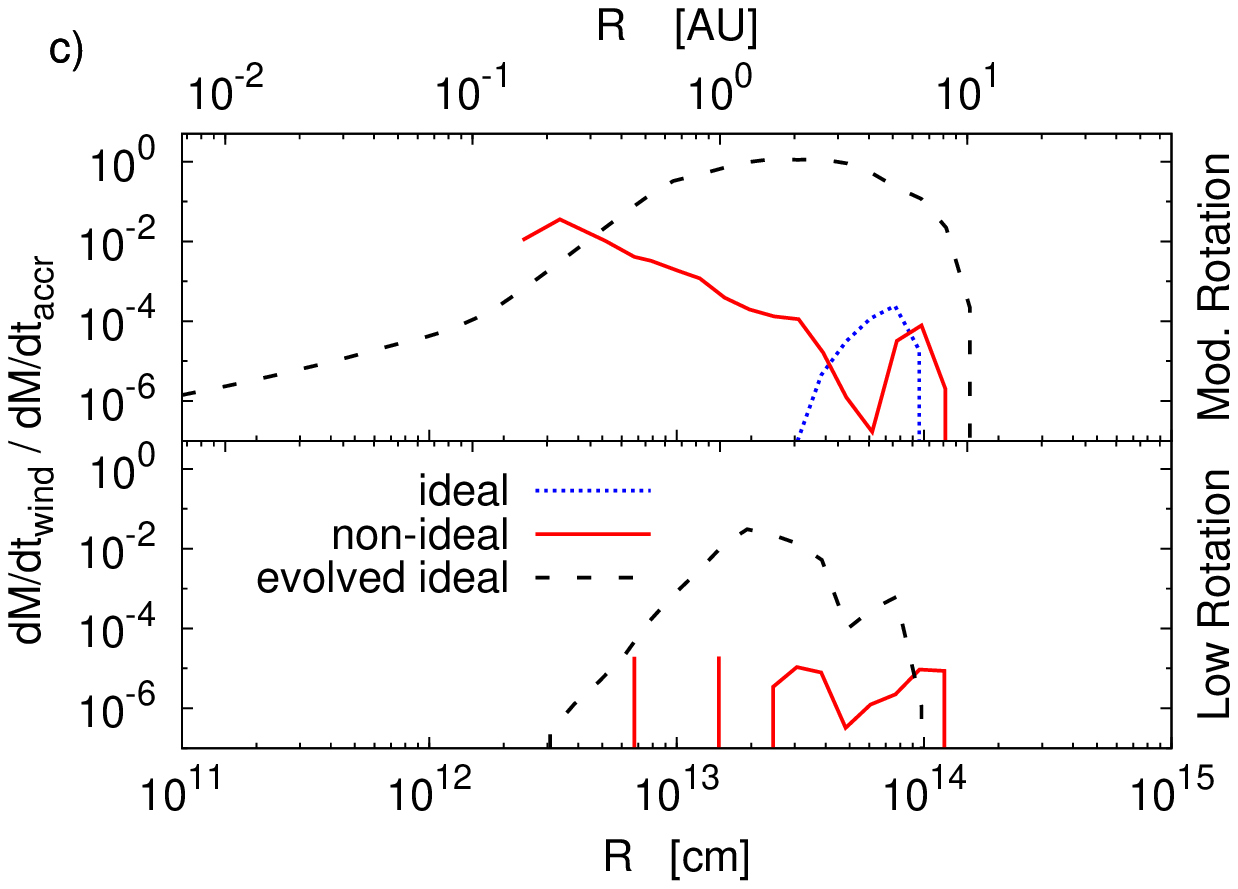}{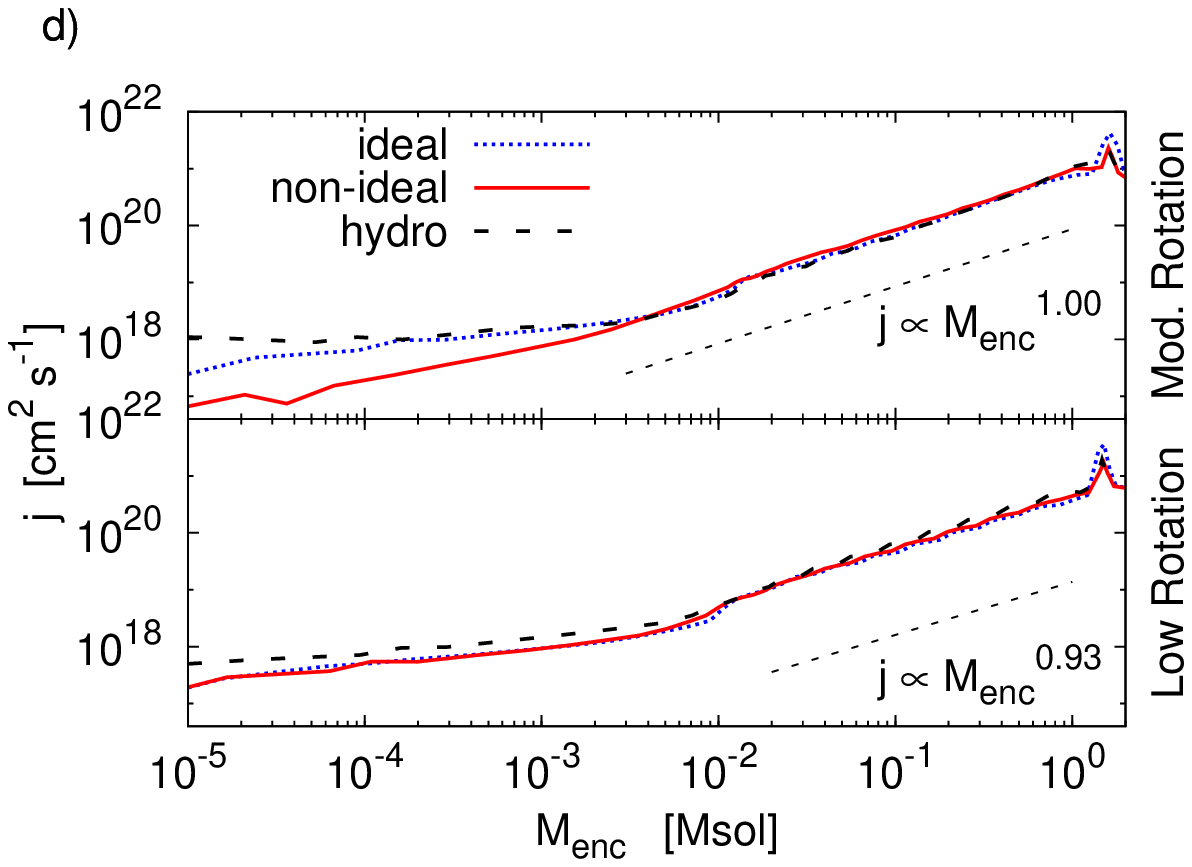}

\plottwo{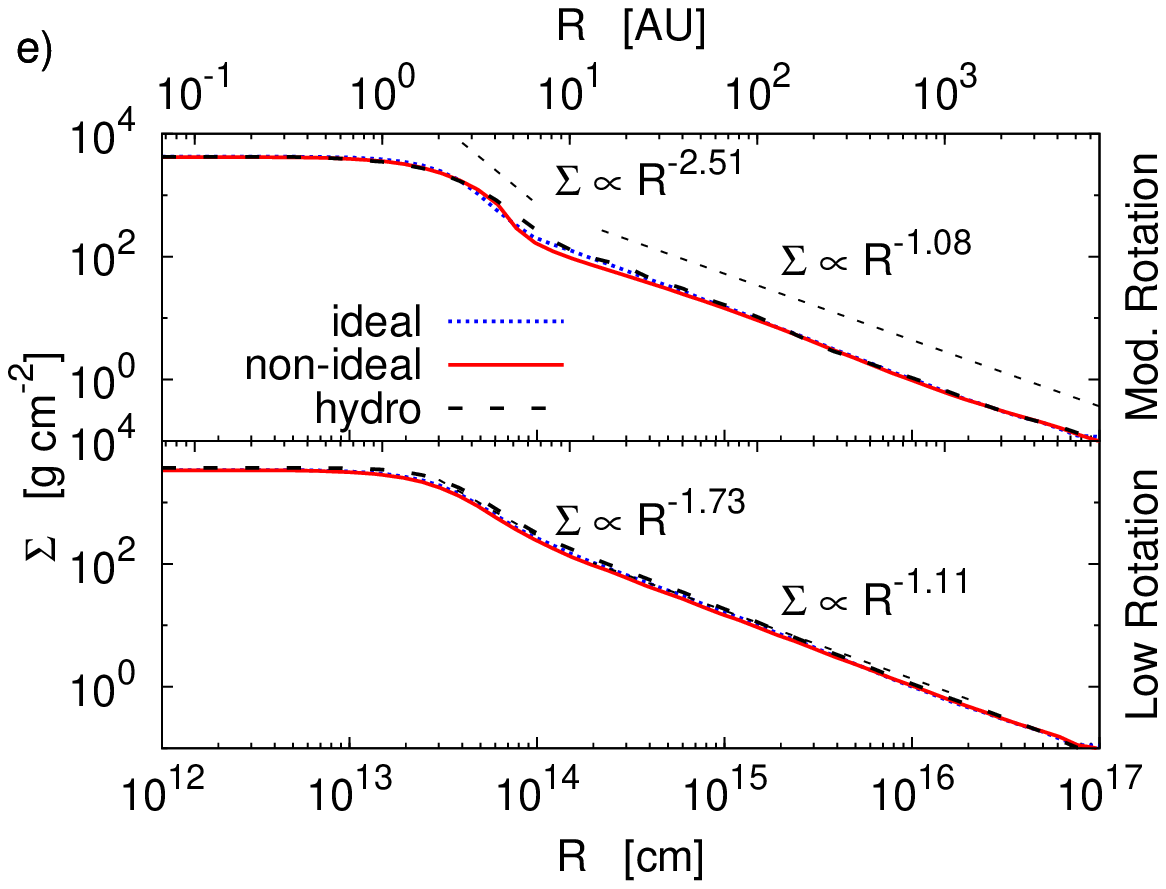}{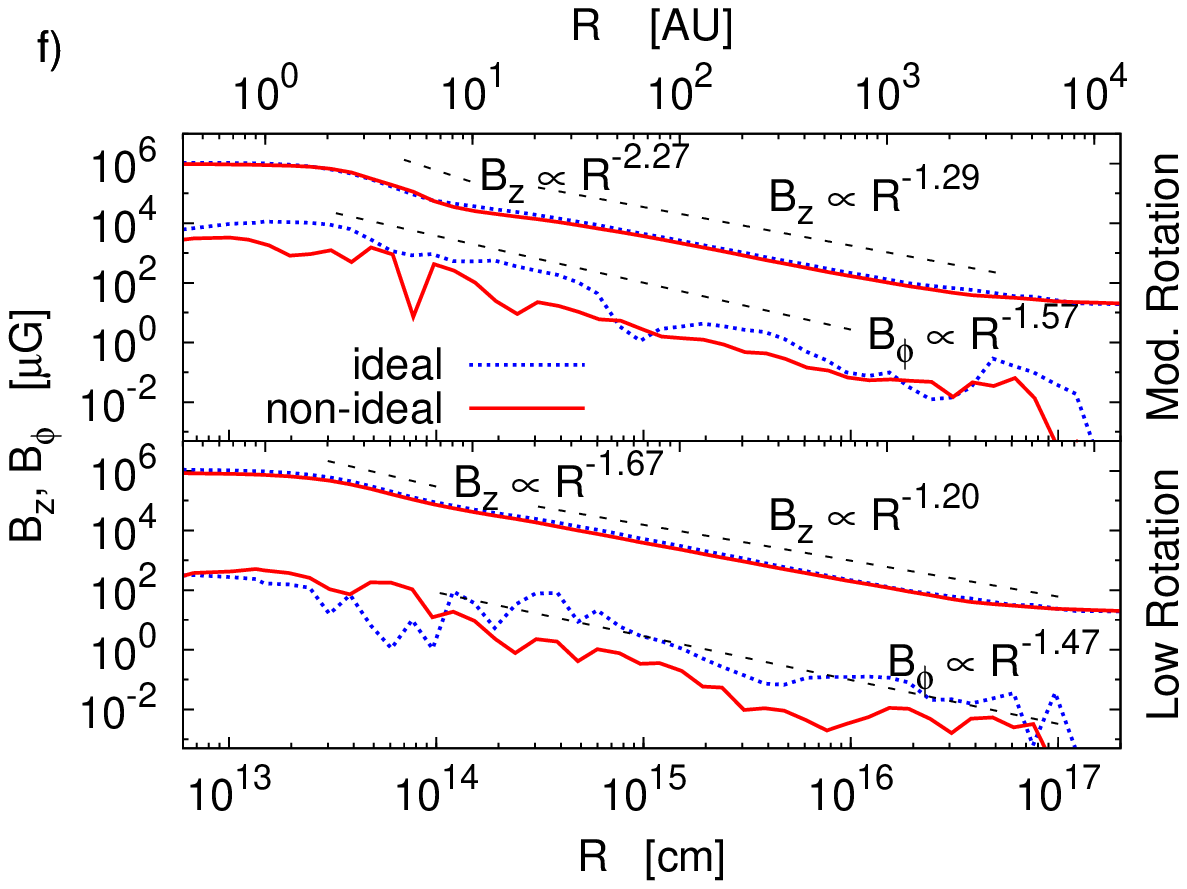}

\caption{\label{fig:2}
Azimuthally averaged plots \citep[density weighted, see ][]{2006ApJ...641..949B} vs. disk radius for ideal MHD, non--ideal MHD, and hydro cases. In all figures the model sets are shown at a common central surface density of $\Sigma_c = 4.2\times10^3~\mathrm{g~cm^{-2}}$ (moderate or mod.~rotation) and $\Sigma_c = 3.4\times10^3~\mathrm{g~cm^{-2}}$ (low rotation), when viewed face on. Panels (a) and (b) show the toroidal velocity in units of infall velocity and Keplerian velocity for the moderate and low rotation model sets, respectively (``evolved" refers to hydro and ideal MHD end--states). The dotted horizontal lines represent a velocity ratio of 1. The dotted vertical lines represent a transition radius of 7 and 6 AU, respectively. Panel (c) shows the outflow efficiency of the magnetized cases as compared to the more evolved ideal MHD case. Panel (d) shows specific angular momentum plotted against enclosed mass. The surface density distributions are shown in panel (e), while the magnetic field distributions are shown in panel (f).
}
\end{figure}

\begin{figure}
\plottwo{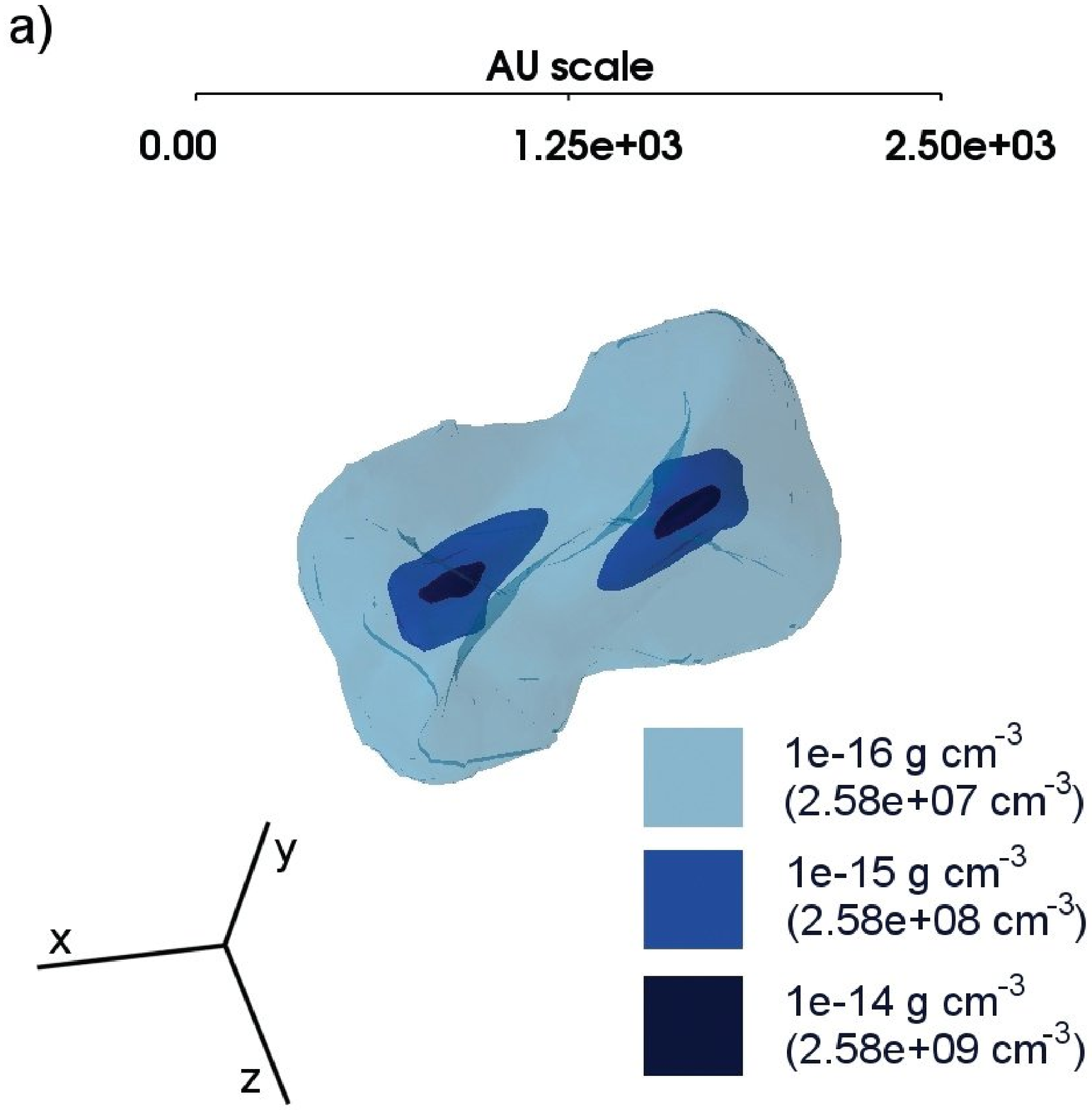}{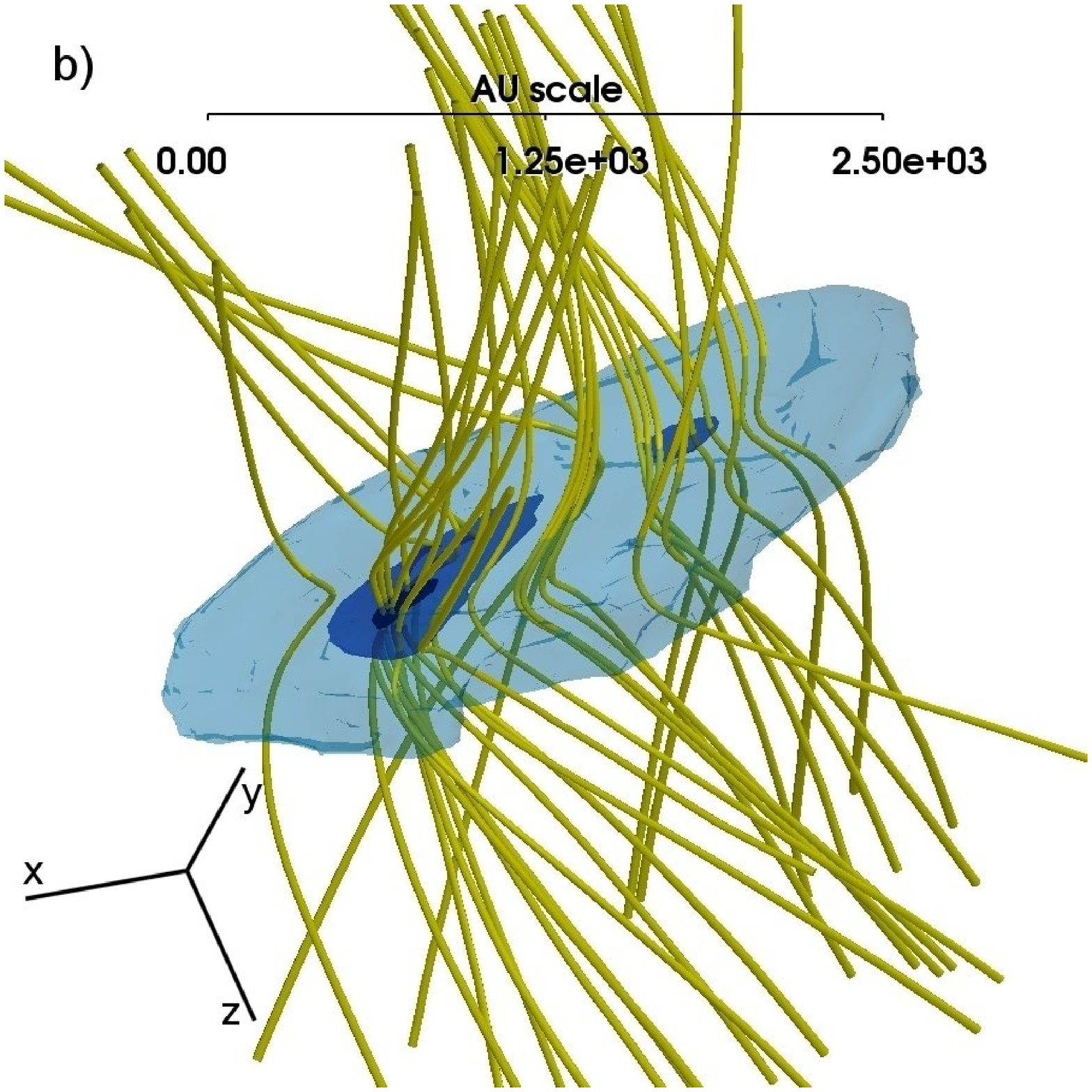}

\epsscale{.45}

\plotone{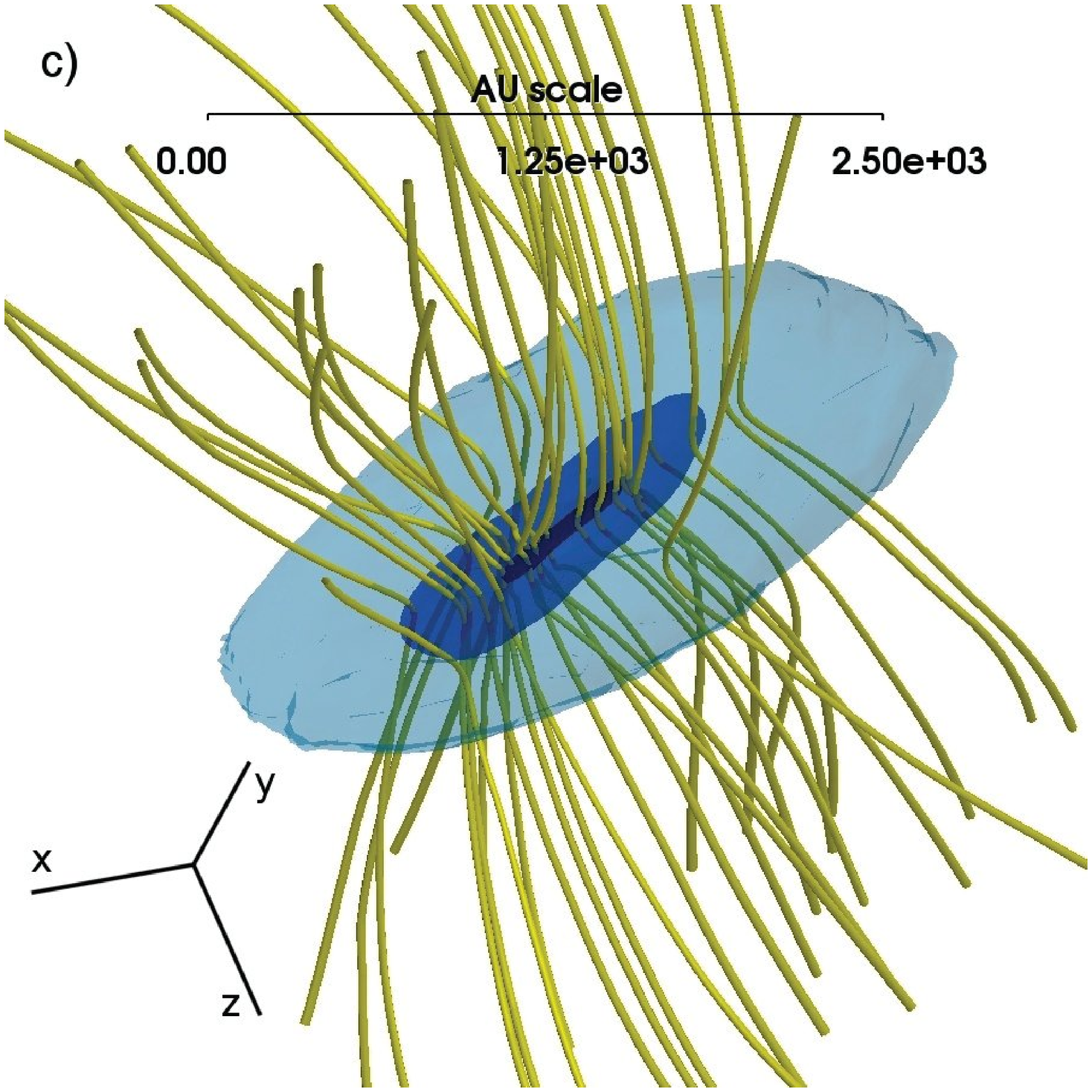}

\caption{\label{fig:3}
Large scale fragmentation of the high rotation model set for (a) hydrodynamic, 
(b) ambipolar diffusion, and (c) ideal MHD cases at similar scales and 
orientations.  Density contours are given in the legend, and magnetic field 
lines are drawn as yellow tubes.}
\end{figure}

Figures \ref{fig:1}(b)-(d) are snapshots of the structure of outflows that are launched
at these early times.  A major result, shown by comparing Figures \ref{fig:1}(b) and (c), is 
that outflows occur even when ambipolar diffusion is active.  
Outflows in a non--ideal MHD collapse have been demonstrated in two-dimensional ambipolar diffusion without drift heating \citep{2009ApJ...698..922M}
and in three-dimensional using an ohmic diffusion approximation \citep{2007ApJ...670.1198M}.  
Our analytical understanding of the early outflow 
mechanism is the magnetic tower \citep{2003MNRAS.341.1360L}.  In this picture, 
the toroidal field in the disk is generated through the rotating flow and 
confined there by the accretion ram pressure.  As the field winds up, it reaches a 
critical strength in which the toroidal pressure is strong enough to overcome the ram 
pressure, moving wrapped field lines away from the disk and taking perfectly 
coupled gas with them -- starting with the initial release in Figure 
\ref{fig:1}(c) and the more developed magnetic bubble shown in Figure 
\ref{fig:1}(d).  This acts to torque down the disk, removing angular 
momentum and increasing accretion rates\footnote{A movie of the development of Figure \ref{fig:1}(d) from Figure \ref{fig:1}(c) is provided online.}.  
In the ambipolar diffusion picture, 
these field lines can seep through the ram pressure as they are not so tightly coupled
to the accreting gas. 
  
We have also discovered the existence of  a centralized, high--speed  component to the ambipolar 
diffusion case's outflow ($v>0.4$ km s$^{-1}$ in an region roughly 1 AU in size).  This component  resembles a jet that is commonly
observed in the evolved perfectly coupled collapse\footnote{The velocity structure of the perfectly coupled outflow is seen in our online movie of the evolution of Figure \ref{fig:1}(d)}.  
We note that the base of this component to the outflow lies on a cushion of 
heat generated by the friction between ions and neutrals as toroidal field lines seep through the accreting gas.  This spike in drift 
heating is typical in ambipolar diffusion mediated C-shocks, regions where 
ion--neutral friction is strong \citep{1991MNRAS.251..119W}.  This local heating 
effect produces strong gradients in temperature with $T>110$ K in regions roughly 5 AU in size, both above and below the midplane.  Within these regions are features roughly 1 AU in size with $T>1000$ K 
(not unheard of in C--shocks).  By comparison, the perfectly coupled outflow 
has a sharp temperature shock at about 50 K leading its outflow.  In contrast, the low rotation model set shows very little drift heating
and a similar weak outflow (Figure \ref{fig:2}(c).  While the vertical field threading the disk $B_z$ is identical for 
all low and moderate rotation, magnetized cases (Figure \ref{fig:2}(f), the toroidal field $B_\phi$ in 
the low rotation, magnetized cases is nearly 2 orders of magnitude smaller than in the
moderate rotation, magnetized cases. 

In Figure \ref{fig:2}(c) we plot the ratio of 
mass loss rates $\dot{M}_\mathrm{wind}$ to mass accretion rates 
$\dot{M}_\mathrm{accr}$ for the moderate and low rotation model sets. For comparison we also plot this ratio for the 
evolved outflow of the perfectly coupled case at later times (labeled `evolved ideal').  
The evolved outflow is very efficient in the moderate rotation, ideal MHD case -- the central outflow component exceeds speeds of 5 km s$^{-1}$ and observed outflow rates of $\dot{M}_\mathrm{wind}/\dot{M}_\mathrm{accr}\sim0.1$ \citep{1995AJ....109.1846H}.  Unfortunately, 
we cannot fully develop the ambipolar outflow much beyond $10^5$ yr due to time step constraints. 
However, from this early footprint, we are led to believe that the outflow 
mechanism is working.
In the low rotation, ideal MHD case, the outflow rates are significantly diminished by nearly 2 orders of magnitude in comparison to mass accretion rates, less than what is typically observed.
This may have consequences on feedback effects in star forming clusters and the assembly of massive stars \citep{2009arXiv0908.4129W}; protostellar energy injection is dependent on the rotational energy of the star forming core.

%
%
\section{Fragmentation and binaries}  

In supercritical cores, three-dimensional ideal MHD calculations have shown that strong perturbations (on the order of 50\%) are needed in order for cores with $\Gamma\sim2$ to fragment and form binaries, even with high rotation \citep{2007MNRAS.377...77P, 2008A&A...477...25H}.
We find that none of the low and moderate rotation cases, magnetized or not, fragment.
We therefore ran a high rotation model set with four times the angular rotation than our earlier model 
set (roughly 20 times the rotational energy, an extreme rotation 
 \citep{2007MNRAS.382...73T}), similar to the simulations of \citet{2007MNRAS.377...77P} with a 10\% perturbation amplitude.  In Figure \ref{fig:3} we plot three-dimensional density contours 
for all three cases of the high rotation model at the largest common central 
column density ($\Sigma_c = $~68 g cm$^{-2}$).   The hydrodynamic case quickly 
fragments
into a binary with a wide separation of roughly 1000 AU.  The perfectly magnetized 
case forms a bar and does not fragment, stabilized through its magnetic pressure 
as well as through the effects of magnetic braking.  Meanwhile, the ambipolar
diffusion case produces an intermediate result.  Early on a ring is formed which 
fragments into a bar--like object with a small companion at about 1000 AU. 

For sufficiently high rotation ambipolar diffusion allows binaries to form without the need of strong (e.g.~$\gtrsim10\%$ in amplitude) perturbations.  For moderate rotation rates however, fragmentation can take place only if large perturbation amplitudes ($\approx 50$\%) are used in the initial conditions.  Such a perturbation would amount to a 1\% change in our BE sphere's gravitational binding energy.  This relatively low level of `turbulent' energy has been measured in B68 \citep{2003ApJ...586..286L}. This would suggest that a low level of turbulent energy (e.g.~a few percent of the gravitational energy) is required in order to reproduce the overwhelming tendency for quiescent cores to form binaries in face of strong magnetic support.

\section{Conclusions}
 In the early stages of a purely hydrodynamic collapse of a moderately rotating core, material joins the protostar by accreting through a chaotic series of bars
and spiral waves. Our results show that in magnetized
collapses however, the magnetic field suppresses these wave modes, and a small, regular disk appears at the earliest times.  
The weakening of magnetic control by ambipolar diffusion is insufficient to guarantee the formation of binary stars in typical cores with moderate rotation. Our results suggest that modest turbulent amplitudes ($>$ 10\%) appear to be required.
Regarding planet formation, the central density structure of early disks (Figure \ref{fig:2}(e) falls off as $\Sigma \propto r^{-(1.7-2.5)}$ much more quickly with disk radius than do protoplanetary disk models at later times, wherein $\Sigma \propto r^{-1.5}$ or $r^{-1}$.  
 As the collapse winds up the field early outflows appear -- even for partially coupled disks --  and feed
  angular momentum and mechanical energy back into the star forming neighborhood.
The density and magnetic structure of ideal and partially coupled disks are quite similar -- the main difference
is in the strength of the wound up magnetic field (at least an order of magnitude weaker in the ambipolar diffusions case).

Finally, we note that evidence for early localized intense drift heating near the disk at the accretion shock, which heats materials
up to 1,000K and beyond in a localized region, may be preserved in the observed composition of comets and meteorites.  
The rapid heating, and subsequent cooling of those crystalline Mg--rich silicate materials that
passed through the accretion shock and through the region of high ambipolar heating is reminiscent of heating events that must have occurred
for some of these materials seen in cometary grains \citep{2007prpl.conf..815W}.  These events have been attributed to shock heating by spiral waves out to 10 AU in disks \citep{2005ASPC..341..849D}.
As seen in Figure 1(b), only a 
portion of the accreted disk material will pass through this localized heated region. Subsequent 
radial turbulent mixing of this flash heated material with the bulk of material that passed through a more gentle heating environment could in principle
contribute to the wide mixture of thermal histories preserved in cometary grain materials.


\section{Acknowledgements}
We thank Robi Banerjee, Eve~C.~Ostriker, and James Wadsley for fruitful discussions. 
D.D. was supported by McMaster University and R.E.P.~ by the Natural Sciences and Engineering
Research Council of Canada. We are pleased to acknowledge the SHARCNET HPC Consortium for the use of its facilities
at McMaster University.
The software used in this work was in part developed by the DOE--supported ASC/Alliances Center for Astrophysical Thermonuclear Flashes at the University of Chicago.
This research was supported in part by the National Science  
Foundation under
grant PHY05--51164.

\bibliographystyle{apj} 
\bibliography{/home/duffindf-roam/school/papers-2008-laptop/master}

\begin{table}
\centering
\begin{tabular}{l|rrr}
\hline
\hline
Model Set  & Low Rotation &  Moderate Rotation & High Rotation \\
\hline 
$\Omega t_\mathrm{ff}=\Omega \sqrt{3\pi/32 G \rho_0}$ & $0.1$  &  $0.3$ & $1.2$ \\
Angular rotation ($\Omega$, s$^{-1}$) & $1.18\times10^{-13} $ &  $3.52\times10^{-13}$ & $1.41\times10^{-12}$ \\
Rotational beta ($\beta_\mathrm{rot}$) & $0.0052$ & $0.046$ & $0.74$\\
\end{tabular}
\caption{\label{tab:1}
Model parameters. Each model set has a mass of $M_\mathrm{core}=1.18 M_\odot$, core temperature of $T_\mathrm{core}=20~\mathrm{K}$ ($c_{s_\mathrm{core}} = 0.267~\mathrm{km~s^{-1}}$), external temperature of $T_\mathrm{ext}=200~\mathrm{K}$ ($c_{s_\mathrm{ext}} = 0.845~\mathrm{km~s^{-1}}$), $\Gamma = 3.5$, and $\beta = 46.01$ and is run in three cases: hydrodynamic, ideal MHD, and ambipolar diffusion. 
}
\end{table}

\end{document}